# The end-to-end simulator of the ATHENA X-IFU Cryogenic AntiCoincidence detector (CryoAC)

M. D'Andrea*[a], C. Macculi [a], S. Lotti [a], L. Piro [a], A. Argan[b], G. Minervini[b], G. Torrioli[c], F. Chiarello[b], L. Ferrari Barusso[d], F. Gatti[d,e], M. Rigano[d,e]

[a] INAF/IAPS, Via del Fosso del Cavaliere 100, 00133 Rome, Italy;
[b] INAF Headquarters, Viale del Parco Mellini 84, 00136 Rome, Italy;
[c] CNR/IFN Roma, Via del Fosso del Cavaliere 100, 00133 Rome, Italy;
[d] University of Genoa, Via Dodecaneso 33, 16146 Genoa, Italy;
[e] INFN Genoa, Via Dodecaneso 33, 16146 Genoa, Italy.

*mail: matteo.dandrea@inaf.it

## ABSTRACT

The X-IFU is one of the two instruments of ATHENA, the next ESA large X-ray observatory. It is a cryogenic spectrometer based on an array of TES microcalorimeters. To reduce the particle background, the TES array works in combination with a Cryogenic AntiCoincidence detector (CryoAC). The CryoAC is a 4-pixel detector, based on ~1 $cm^2$ silicon absorbers sensed by Ir/Au TES. It is required to have a wide energy bandwidth (from 20 keV to ~1 MeV), high efficiency (< 0.014% missed particles), low dead-time (< 1%) and good time-tagging accuracy (10 μs at 1σ). An end-to-end simulator of the CryoAC detector has been developed both for design and performance assessment, consisting of several modules. First, the in-flight flux of background particles is evaluated by Geant4 simulations. Then, the current flow in the TES is evaluated by solving the electro-thermal equations of microcalorimeters, and the detector output signal is generated by simulating the SQUID FLL dynamics. Finally, the output is analyzed by a high-efficiency trigger algorithm, producing the simulated CryoAC telemetry. Here, we present in detail this end-to-end simulator, and how we are using it to define the new CryoAC baseline configuration in the new Athena context.

# 1. INTRODUCTION

ATHENA is the next ESA flagship X-ray observatory (0.2-12 keV) [1]. The mission has just successfully completed a redefinition process to meet the cost-cap set by ESA, and the launch of “new Athena” is planned for the end of 2030. The X-IFU is one of the two instruments on board [2]. It is a cryogenic spectrometer based on a large array of Transition Edge Sensor (TES) microcalorimeters. The TES array works in combination with a Cryogenic AntiCoincidence detector (CryoAC) [3], aimed to reduce the particle background of the instrument and thus enabling the observation of faint and distant sources.

The CryoAC is a 4-pixel detector, each pixel based on ~1 cm$^2$ silicon absorbers sensed by Ir/Au TES. Each pixel is readout by a DC-SQUID operating in Flux Locked Loop (FLL) configuration, and it features a heater deposited on the absorber for diagnostic and calibration purposes. To sense the energy deposition of background particles in the absorbers, distributed around the most probable value of ~ 150 keV expected for MIP (Minimum Ionizing Particle), the CryoAC is required to have wide energy bandwidth, with a low energy threshold < 20 keV and a saturation energy > 1 MeV. Furthermore, it shall have high efficiency (< 0.014% of missed particles above 20 keV), low dead-time (< 1%) and good time-tagging accuracy (10 μs at 1σ). Finally, it shall ensure thermal and electromagnetic compatibility with the TES array within the Focal Plane Assembly (FPA). To do this, the CryoAC cold stage thermal design has to be compliant with a heat load budget $P_{TOT}$ < 40 nW at 50 mK, it shall operate with a bias current $I_{BIAS}$ < 3 mA (< 1 mA goal), and its impact on the energy resolution of the TES array shall be less than 0.1 eV (as root-sum-square contribution to the FWHM energy resolution of the X-IFU instrument). To assess this last point, a joint TES array – CryoAC Demonstration Models (DM) test has been successfully performed at chip level in 2020 [4], when the baseline readout for the X-IFU was the Frequency Domain Multiplexing (FDM). The same tests will soon be repeated in Time Division Multiplexing (TDM) readout as part of the FPA DM 1.1 campaign [5].

Recently, we have developed an end-to-end simulator of the CryoAC detector, which consists of several modules (Fig. 1). First, the expected in-flight flux of background particles and the corresponding energy depositions on the absorbers are evaluated by detailed Geant4 simulations, and timed by assuming Poissonian statistics to generate an event list. Then, the current flow in the TES is evaluated time-by-time by solving the electro-thermal equations of microcalorimeters, and the detector output signal is generated by simulating the DC-SQUID FLL dynamics. Finally, the output is analyzed by a custom high-efficiency trigger algorithm, producing the simulated CryoAC telemetry.

We use this simulator to drive the CryoAC design, by tuning the detector parameters to meet the mission requirements. Here, we present in detail the CryoAC end-to-end simulator, and how we used it to define the new baseline CryoAC configuration in the new Athena context.

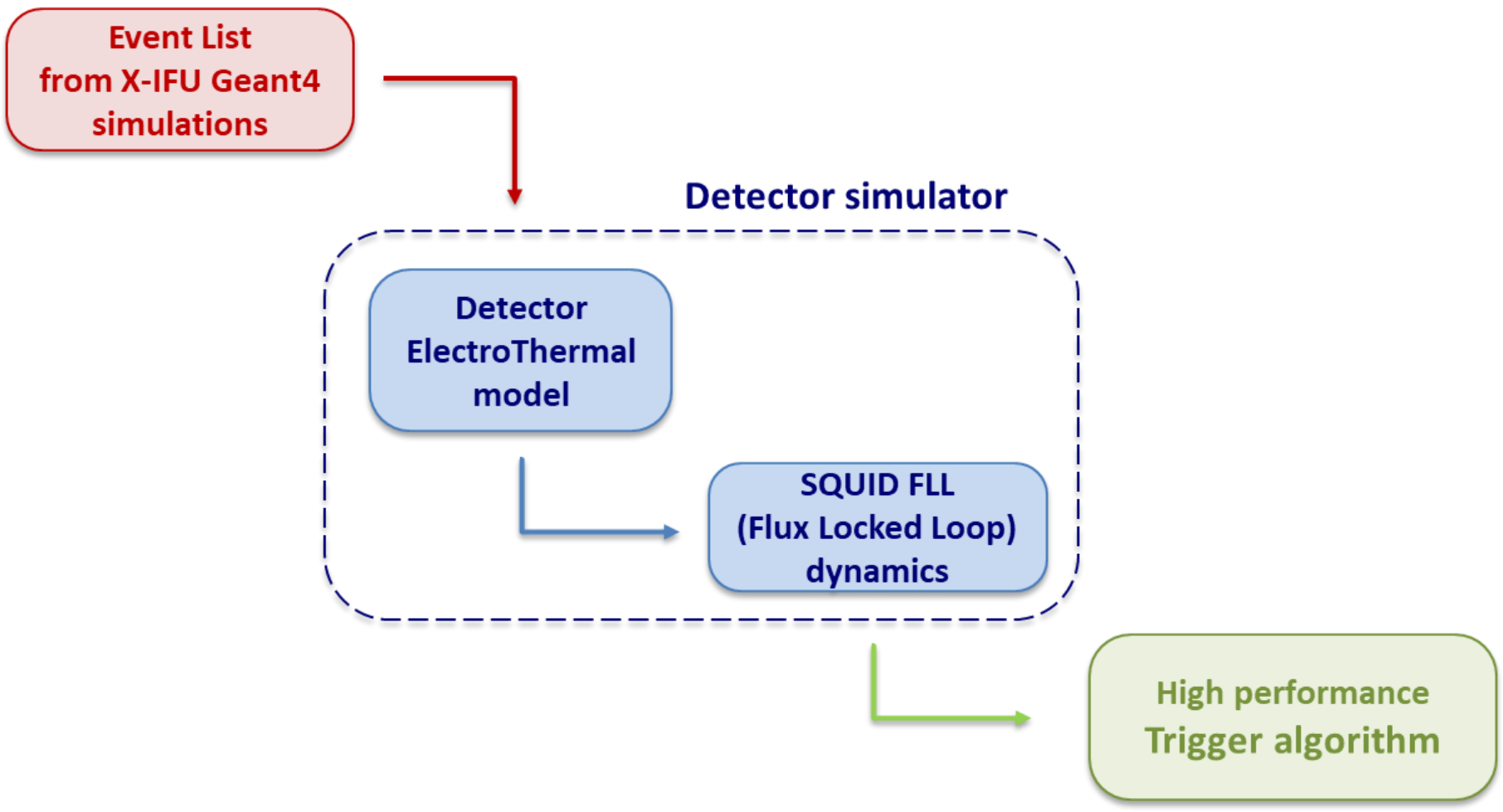

Figure 1. Schematics of the CryoAC end-to-end simulator structure.

## 2. THE EVENT LIST

To simulate the CryoAC behaviour in the L1 spacecraft environment, we have generated an event list containing the expected arrival times and energy depositions of the particles interacting with the detector.

The expected in-orbit particle flux and the spectrum of the deposited energies have been obtained from the official X-IFU Geant4 simulations, which are developed to assess the instrument residual background (see [6] for details). They take into account the contributions from the primary Galactic Cosmic Rays (GCR protons, alpha particles and electrons) and from the secondary particles generated inside the spacecraft. We have rescaled the flux assuming a reduction of the CryoAC area in new Athena from 4.9 $cm^2$ to ~ 3.0 $cm^2$ (i.e. from 1.23 $cm^2$ to ~ 0.75 $cm^2$ per pixel), due to the reduced Field Of View (FoV) of the instrument. The resulting spectrum of deposited energies is reported in Fig. 2. Arrival times have been then generated assuming Poisson distributions for the different particle species.

The reference CryoAC pixel event list contains 571'135 particles interacting with the detector in 190.8 ks, resulting in a count rate of ~ 3.0 cts/s/pixel. Tab. 1 shows the number of particles depositing energy above different thresholds. The maximum deposited energy is ~ 160 MeV.

Note that the CryoAC efficiency requirement (< 0.014% of missed particles above 20 keV) translates into less than 76 missed events above 20 keV in this event list.

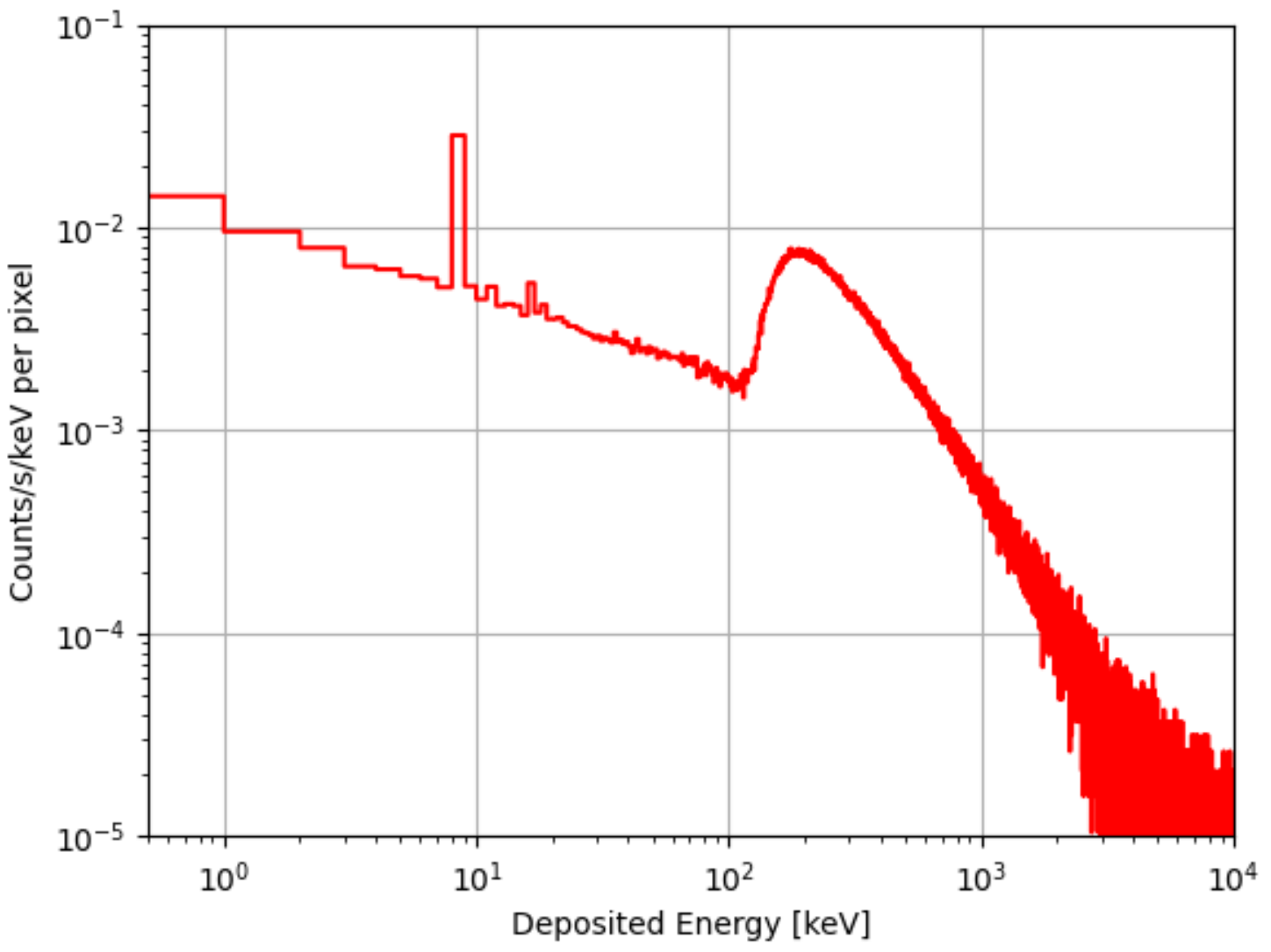

Figure 2. Spectrum of the energy deposition on the CryoAC expected during on-flight operations.

Table 1. Number N of events having an energy deposition higher than the energy E in the CryoAC pixel event list. $N_{TOT}$ = 571'135.

| E [keV] | N (> E) | N (> E) / $N_{TOT}$ |
|---|---|---|
| 6 | 561'556 | 0.983 |
| 20 | 544'995 | 0.954 |
| 1'000 | 80'137 | 0.140 |
| 2'000 | 31'804 | 0.056 |
| 6'000 | 6'387 | 0.011 |
| 10'000 | 2'673 | 0.004 |
| 100'000 | 10 | 0.00002 |

## 3. THE DETECTOR SIMULATOR

The core of the CryoAC end-to-end simulator is the detector simulator, which solves time-by-time the equations describing the electro-thermal status of the system and the DC-SQUID Flux Locked Loop (FLL) operations, by using the Euler method. In this section, we first present the detector electro-thermal model and then the SQUID-FLL dynamics.

### 3.1 Detector electro-thermal model

The CryoAC pixel thermal and electrical equivalent circuits are shown in Fig. 3 and Fig. 4.

Note that the model assumes a thermal decoupling between the TES and the absorber. Indeed, at very low temperature, the thermal impedance between the electron and the phonon systems in the TES can be important, and the TES free electrons have generally a different temperature from both the phonons in the TES itself (which can be neglected in the model due to their very small heat capacity) and the phonons in the silicon absorber [7][8].

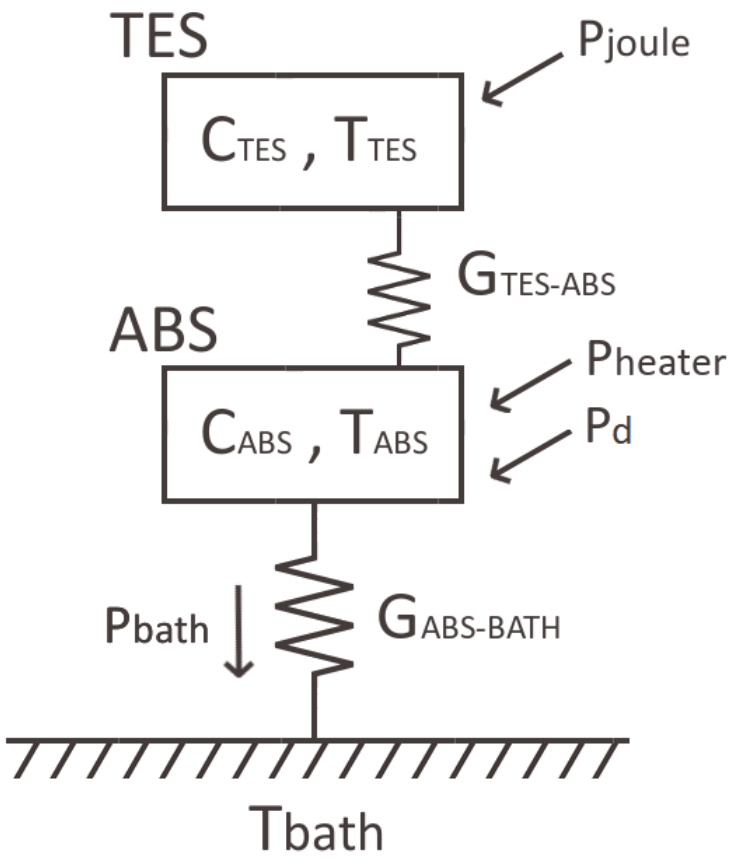

Figure 3. Thermal circuit representing a CryoAC pixel. $C_{TES}$ and $T_{TES}$ are the TES electrons heat capacity and temperature. $P_{joule}$ is the joule power dissipated by the TES bias. The TES is connected to the silicon absorber through the thermal conductance $G_{TES-ABS}$ (total equivalent conductance including TES electron-phonon conductance and the Kapitza interface conductance). $C_{ABS}$ and $T_{ABS}$ are the absorber phonons heat capacity and temperature. $P_{heater}$ is the power dissipated on the absorber by the on-chip heater. $P_d$ is the power dissipated on the absorber via energy deposited by an incoming particle/photon. $G_{ABS-TES}$ is the thermal conductance between the absorber and the thermal bath at $T_{bath}$ temperature, through which flows the total power $P_{bath} = P_{joule} + P_{heater} + P_d$. From [7].

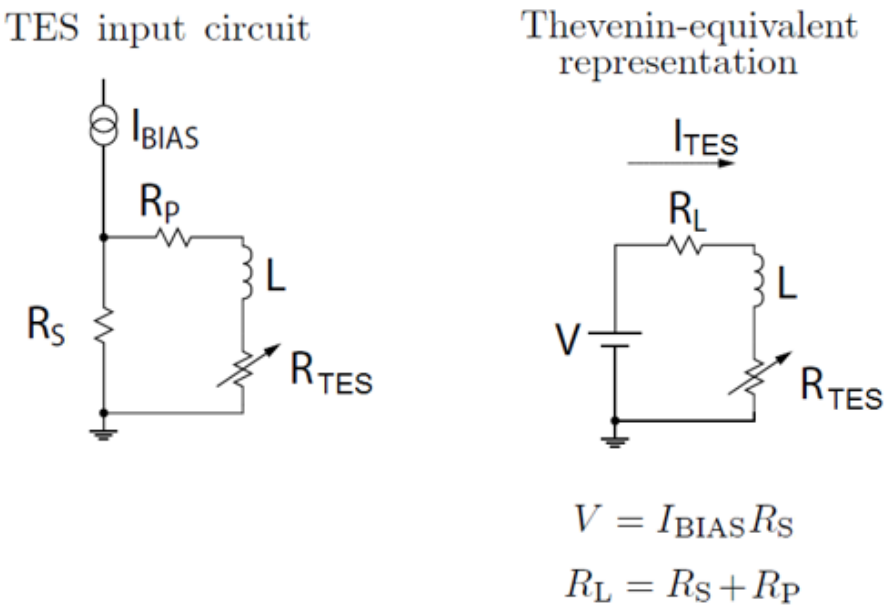

Figure 4. The TES electrical circuit and its Thevenin-equivalent representation. (Left) The bias current $I_{BIAS}$ is applied to the shunt resistor $R_S$ in parallel with the TES, the total inductance L (including the input coil and the stray wire contribution) and a parasitic resistance $R_P$. (Right) In the Thevenin- equivalent circuit a bias voltage $V = I_{BIAS} R_S$ is applied to the series of a load resistor $R_L = R_S + R_P$, the TES and the input coil, through which the $I_{TES}$ current is flowing. From [8].

The system evolution is described by the following set of three differential equations:

$$C_{TES}\frac{dT_{TES}(t)}{dt} + P_{TES-ABS} = P_{joule} \tag{1}$$

$$C_{ABS}\frac{dT_{ABS}(t)}{dt} + P_{bath} = P_{TES-ABS} + P_{joule} + P_{Heater} + P_D \tag{2}$$

$$L\frac{dI_{TES}(t)}{dt} + I_{TES}(t)(R_{TES} + R_L) = V \tag{3}$$

where:

$$P_{joule} = I_{TES}{}^2 R_{TES} = I_{TES}{}^2 R_{TES}(T_{TES}(t)) \tag{4}$$

$$P_{TES-ABS} = k_{TA}(T_{TES}(t)^{n_{TA}} - T_{ABS}(t)^{n_{TA}}) \tag{5}$$

$$P_{bath} = k_{AB}(T_{ABS}(t)^{n_{AB}} - T_B(t)^{n_{AB}}) \tag{6}$$

$$P_d = E_d\, \delta(t) \tag{7}$$

$R_{TES}(T_{TES}(t))$ is the in-time description of the TES transition curve and $E_d$ is the energy deposited on the absorber by an incoming particle/photon.

In the simulator, we translate the set of equations in discrete numerical approximation, and we add the noise contributions ($N_{PHON,TES}$ is the phonon noise at the TES-absorber interface, $N_{PHON,ABS}$ is the phonon noise at the absorber-bath interface, $N_{TES,J}$ is the TES Johnson noise, $N_{SHUNT,J}$ is the shunt resistor Johnson noise):

$$T_{TES\,(i+1)} = T_{TES\,(i)} + dX \cdot \left[I_{TES\,(i)}{}^2 R_{TES\,(i)} - k_{TA}\left(T_{TES\,(i)}{}^{n_{TA}} - T_{ABS\,(i)}{}^{n_{TA}}\right) + N_{PHON,TES}\right] / C_{TES\,(i)} \tag{8}$$

$$\begin{aligned} T_{ABS\,(i+1)} &= T_{ABS\,(i)} + E_D/C_{ABS\,(i)} + \\ &dX \cdot \left[k_{TA}\left(T_{TES\,(i)}{}^{n_{TA}} - T_{ABS\,(i)}{}^{n_{TA}}\right) - k_{AB}\left(T_{ABS\,(i)}{}^{n_{AB}} - T_{B\,(i)}{}^{n_{AB}}\right) + P_{Heater} + N_{PHON,ABS}\right] / C_{ABS\,(i)} \end{aligned} \tag{9}$$

$$I_{TES\,(i+1)} = I_{TES\,(i)} + dX \cdot \left[V - I_{TES\,(i)}\left(R_{TES\,(i)} + R_L\right) + N_{TES,J} + N_{SHUNT,J}\right] / L \tag{10}$$

$dX$ is the discrete time step adopted in the simulation. Noise contributions are randomly generated time-by-time by assuming Gaussian statistics and the following RMS value [7]:

$$N_{TES,J\,(RMS)} = \sqrt{4K_B T_0 (R_0 + R_P)/(2 \cdot dX)} \tag{11}$$

$$N_{SHUNT,J\,(RMS)} = \sqrt{4K_B T_S R_S/(2 \cdot dX)} \tag{12}$$

$$N_{PHON,ABS\,(RMS)} = \sqrt{4n_{AB}/(2n_{AB}+1) \cdot K_B T_{ABS,0}{}^2/G_{AB,0} \cdot 1/(2 \cdot dX)} \tag{13}$$

$$N_{PHON,TES\,(RMS)} = \sqrt{4n_{TA}/(2n_{TA}+1) \cdot K_B T_{TES,0}{}^2/G_{TA,0} \cdot 1/(2 \cdot dX)} \tag{14}$$

where $G_{AB,0} = n_{AB} k_{AB} T_{ABS,0}{}^{(n_{AB}-1)}$, $G_{TA,0} = n_{TA} k_{TA} T_{TES,0}{}^{(n_{TA}-1)}$ and $T_S$ is the temperature of the shunt resistor.

### 3.2 SQUID FLL dynamics model

The SQUID FLL dynamics schematics is shown in Fig. 5.

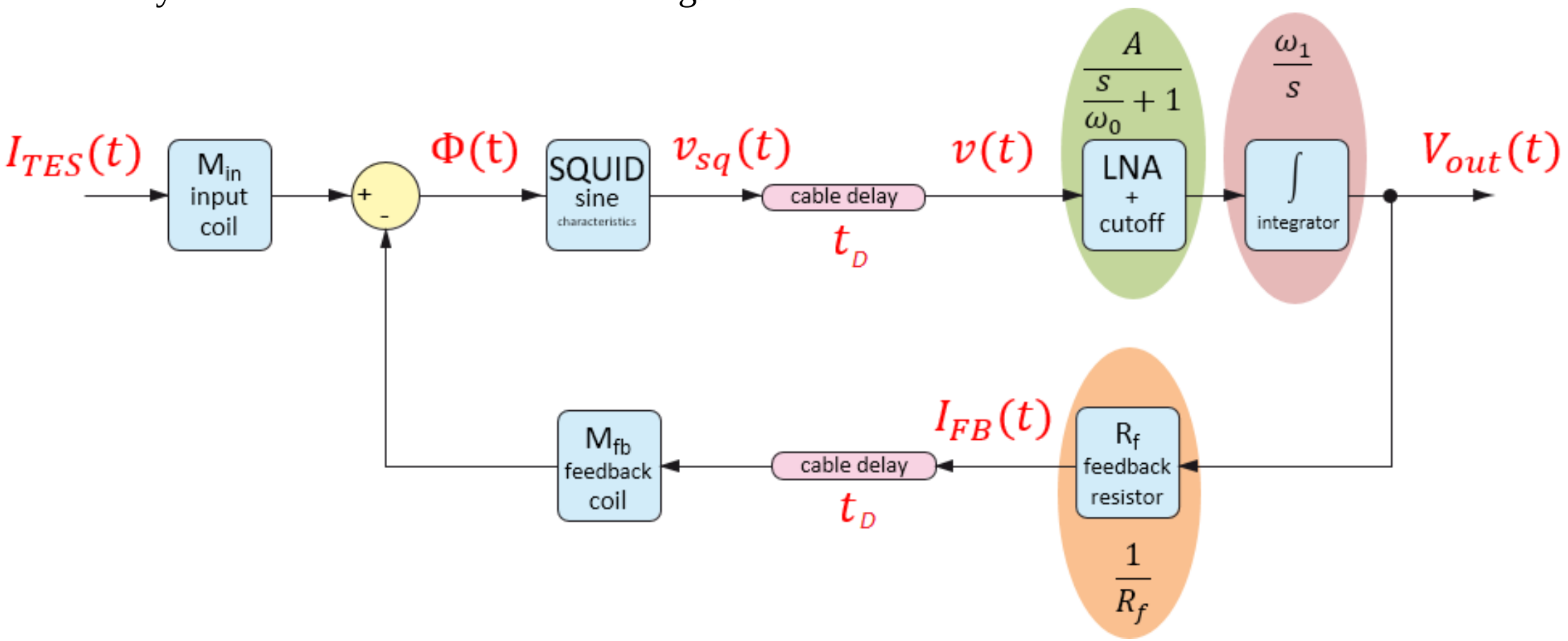

Figure 5. SQUID FLL dynamics schematics. $I_{TES}(t)$ is the current flowing through the input coil, which is coupled to the SQUID by a mutual inductance $M_{in}$. $\Phi(t)$ is the flux concatenated to the SQUID, and $v_{sq}(t)$ the voltage across it. v(t) is the SQUID voltage delayed by the cable between the cold stage and the warm electronics. A is the Low Noise Amplifier (LNA) gain, $\omega_0$ its frequency cutoff and $\omega_1$ is the integrator bandwidth. $I_{FB}$ (t) is the current flowing through the feedback resistor $R_f$, which is delayed by the cable between the warm electronics and the cold stage and coupled to the SQUID by a mutual inductance $M_{fb}$. $V_{out}$ (t) is the output signal of the system.

The system dynamics is described in the frequency domain as follows:

Equation (frequency): $I_{FB}(s) = \frac{1}{Rf}\frac{A}{\left(\frac{s}{\omega_0}+1\right)}\frac{\omega_1}{s}v(s) = \frac{A\omega_0\omega_1}{Rf}\frac{v(s)}{s^2+\omega_0 s}$ (15)

where s is the Laplace variable. The corresponding equation in the time domain is:

Equation (time): $\frac{d^2 I_{FB}(t)}{dt} + \omega_0 \frac{dI_{FB}(t)}{dt} = \frac{A\omega_0\omega_1}{R_F}v(t)$ (16)

This is a second-order differential equation, which is equivalent to a system of two first-order differential equations:

$$J_{FB}(t) = 1/\omega_0 \frac{dI_{FB}(t)}{dt} \quad (17)$$

$$\frac{dJ_{FB}(t)}{dt} + \omega_0{}^2 J_{FB}(t) = \frac{A\omega_0\omega_1}{R_F}v(t) \quad (18)$$

where

$$v(t) = v_{SQ}(t - t_D) = V_{SQUID}\left(\Phi(t - t_D)\right) \quad (19)$$

$$\Phi(t - t_D) = M_{IN}I_{TES}(t - t_D) - M_{FB}I_{FB}(t - t_D - t_D) \quad (20)$$

$V_{SQUID}\,(\Phi)$ is the description of the SQUID characteristics V-Φ curves and $t_D$ the delay induced by the cable between the cold stage and the warm electronics.

In the simulator, we translate the set of equations in discrete numerical approximation and we add the noise contributions ($N_{LNA}$ is the LNA noise, $N_{SQUID}$ is the SQUID noise and $N_{CABLE}$ is the thermal noise of the cables between the cold stage and the warm electronics)$N_{SQUID}$:

$$I_{FB\,(i+1)} = I_{FB\,(i)} + J_{FB\,(i)}\,\omega_0\,dX \tag{21}$$

$$J_{FB\,(i+1)} = J_{FB\,(i)} + \left[-J_{FB\,(i)} + A\omega_1/(\omega_0 R_F)(V_{SQUID(i-i_D)} + N_{LNA} + N_{CABLE})\right]\omega_0\,dX \tag{22}$$

$$\Phi_{(i-i_D)} = M_{IN}\,I_{TES\,(i-i_D)} - M_{FB}\,I_{FB\,(i-i_D-i_D)} + N_{SQUID} \tag{23}$$

where $dX$ is the discrete time step adopted in the simulation. The output signal is finally conditioned by applying a gain factor $G_{WBEE}$ and a low pass filter with cutoff frequency $f_{LP}$, and by introducing the DAQ acquisition noise $N_{DAQ}$:

$$V_{OUT\,(i+1)} = I_{FB\,(i+1)} \cdot \alpha_{WBEE}\,G_{WBEE}\,R_F + V_{OUT\,(i)} \cdot (1 - \alpha_{WBEE}) + N_{DAQ} \tag{24}$$

$$\alpha_{WBEE} = dX/[1/(2\pi f_{LP}) + dX] \tag{25}$$

All the noise contributions are randomly generated time-by-time by assuming Gaussian statistics. The cables thermal noise RMS value is assumed as:

$$N_{CABLE(RMS)} = \sqrt{4K_B T_W R_W/(2 \cdot dX)} \tag{26}$$

where $T_W$ is the average temperature of the wiring and $R_W$ its total resistance.

# 4. SIMULATOR VALIDATION

The simulator has been tuned and validated by the last CryoAC single-pixel prototype, namely DM127, which represents the current reference for a CryoAC pixel. In this section, we show the prototype description inside the simulator and the main results of the simulations. The measurements performed to assess the critical thermal parameters of the system and to validate the simulator are then described in detail in [10].

## 4.1 The CryoAC single-pixel prototype DM127

The CryoAC prototype DM127 is a single-pixel DM-like sample. It is based on a suspended 1 cm$^2$ silicon absorber, sensed by a single Ir/Au TES. It is read-out by a DC-SQUID produced by VTT (model M4A, ret M) and operated by a commercial Magnicon XXF-1 FLL electronics. Fig. 6 shows a picture of the detector assembly. A full description of the sample and the test campaign performed with it is reported in [9].

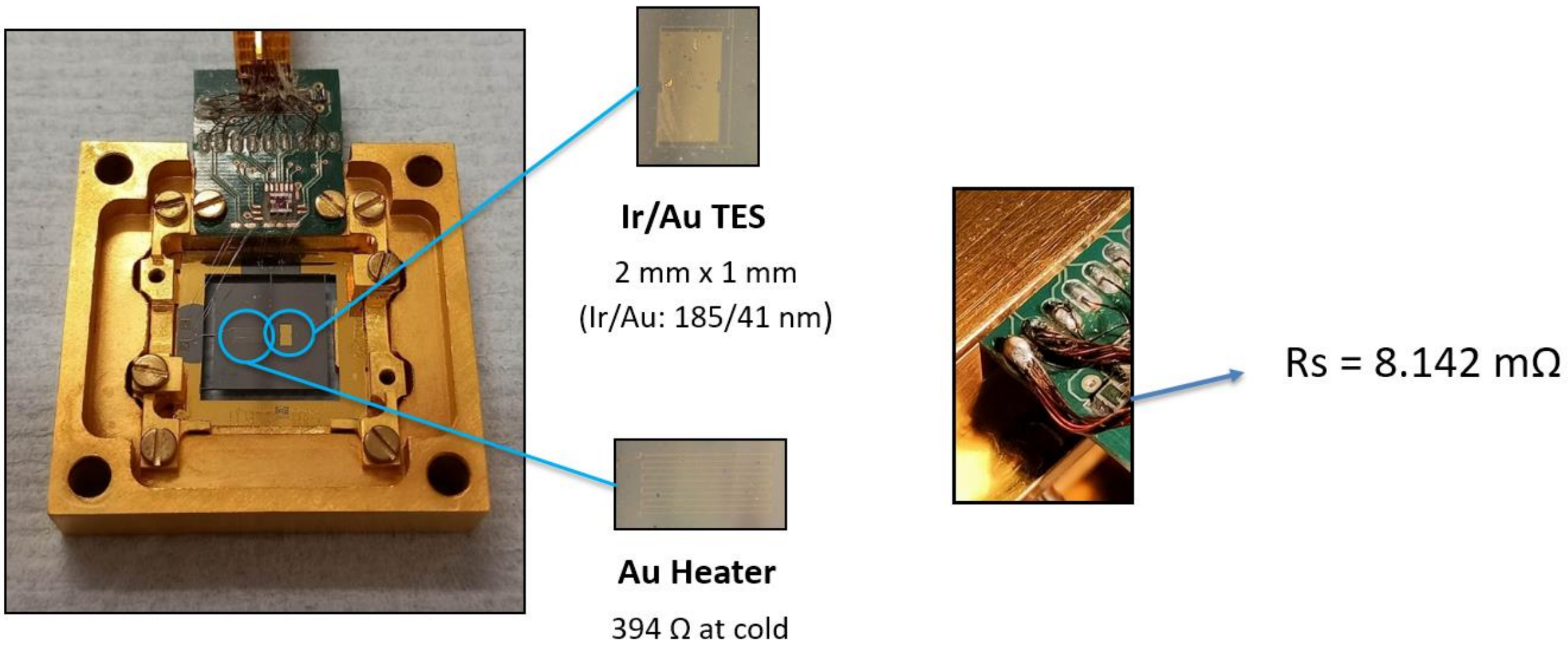

Figure 6. The CryoAC thermal prototype DM#127.

The DM127 R vs T transition curve and its SQUID V-Φ characteristics are reported in Fig. 7 (red points). Inside the simulator, the curves are represented by the following simple functions (blue lines in the figure):

$$R(T) = \frac{R_N}{\left(1 + e^{\left(\frac{\alpha_c}{2}\cdot\left(1-\frac{T}{T_C}\right)\right)}\right)} \tag{27}$$

$$V(\Phi) = V_{SQ,0} + V_{SQ,1} \cdot \sin(2\pi(\Phi - \Phi_0)) \tag{28}$$

where $R_N$ is the normal resistance of the TES, $T_C$ its critical temperature (i.e. the temperature at 50% transition), $\alpha_c$ is the TES thermal responsivity at $T_C$, $V_{SQ,0}$ is the SQUID voltage offset, $V_{SQ,1}$ the V-$\Phi$ amplitude and $\Phi_0$ the SQUID flux offset.

These functions are defined to describe well-shaped curves, but unfortunately, in DM127 both the TES transition and the SQUID characteristics presented issues. Thus, the curves parameters have been chosen to well-describe the curves slope around the TES and SQUID working points (blue arrows in Fig. 7).

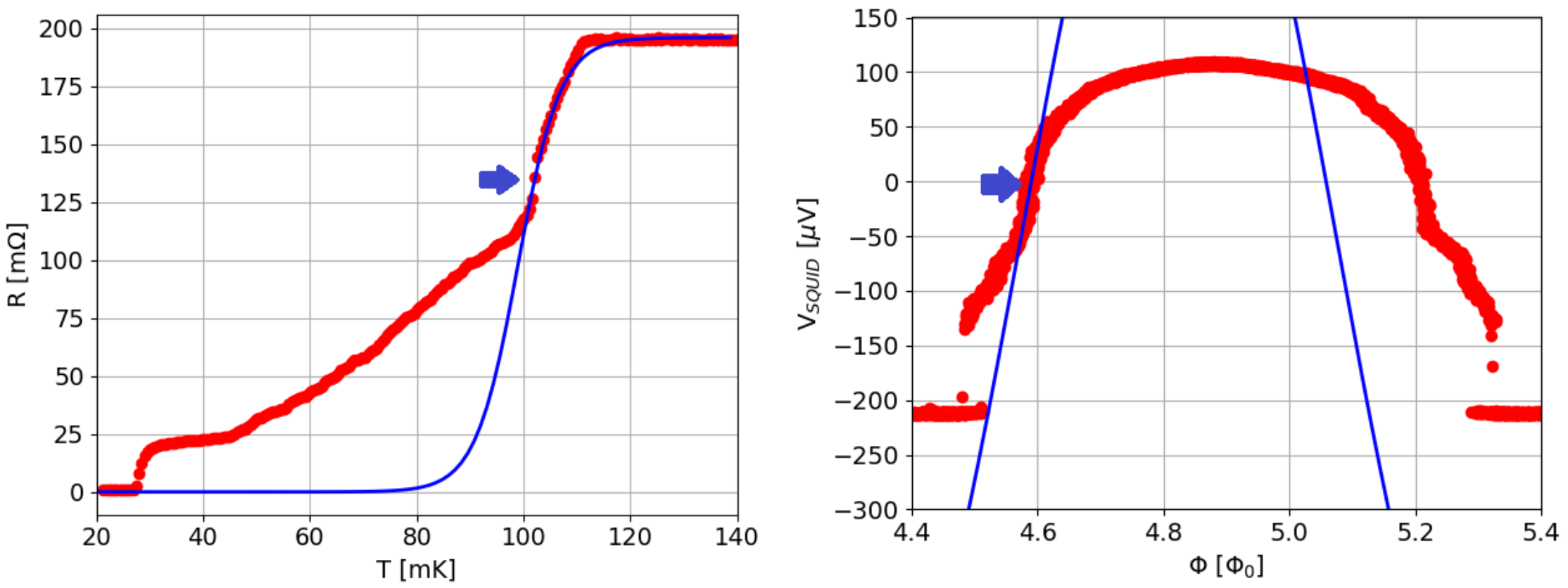

Figure 7. Comparison between the R vs T (Left) and the V-Φ (Right) curves of the DM127 (red points) with their representations inside the simulator (blue lines). The blue arrows represent the TES and SQUID working points.

The parameters used to describe the test environment, the sample thermal characteristics, its bias circuitry and the electronics are reported in Tab. 2.

### 4.2 Comparison between simulator output and real data

We have validated the simulator by comparing its output with real data acquired by CryoAC DM127. This activity is reported in detail in [10]. The simulator is able to describe pulse shape, pulse height and detector noise with an accuracy of a few percent in a wide energy range (from few keV up to the saturation regime above ~ 5 MeV). As example, Fig. 8 shows the comparison between a 6 keV pulse acquired by DM127 illuminated by an $^{55}Fe$ source (blue line), and the output of the simulator assuming a 6 keV energy deposition (red line). We consider this accuracy fully sufficient for the aims of the simulator.

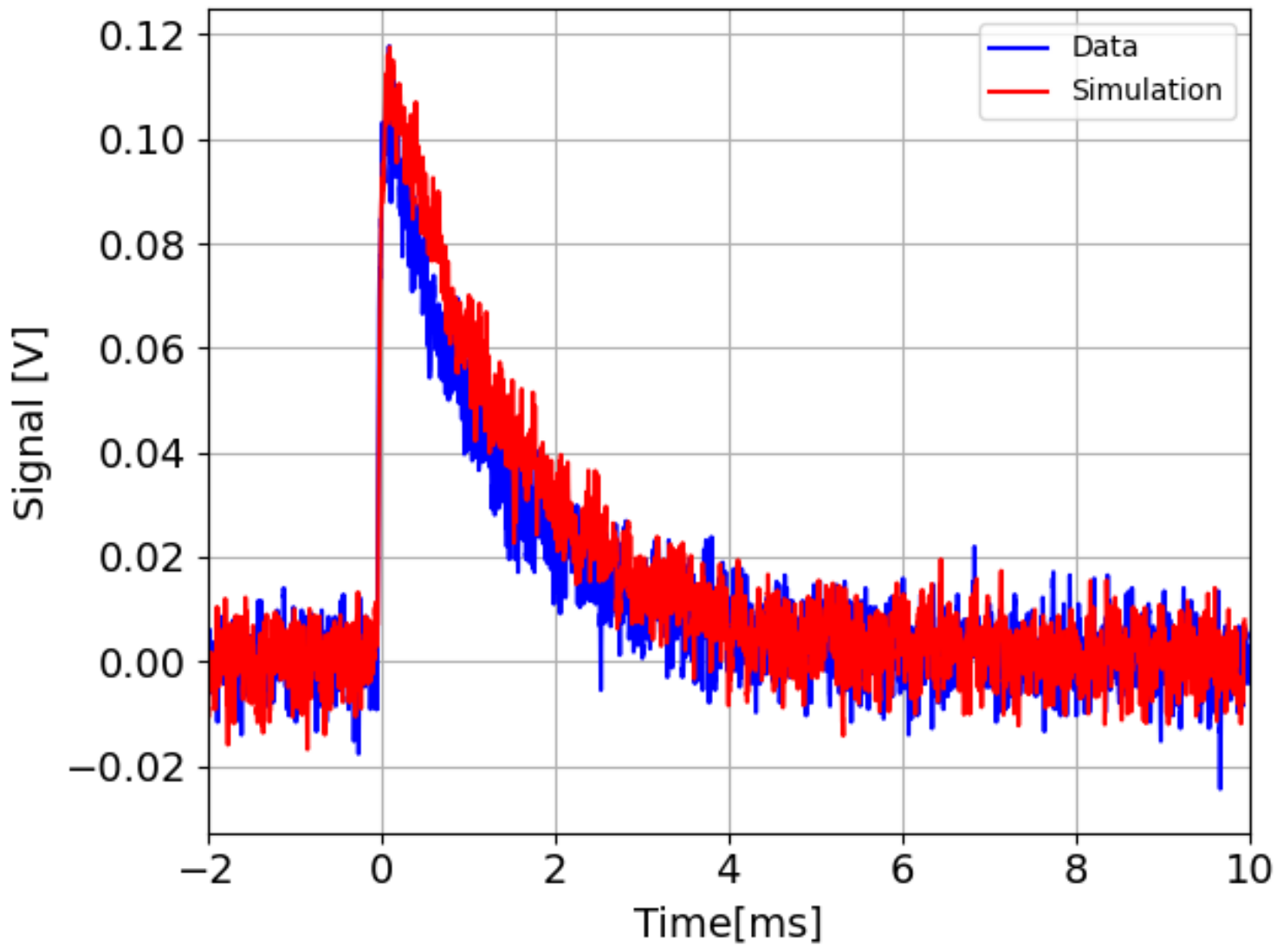

Figure 8. Comparison between a 6 keV pulse acquired by the CryoAC DM127 (blue line) and the output of the simulator with a 6 keV energy deposition (blue line).

Table 2. Parameters used to describe the DM127 operations: “DM127 (1 $cm^2$) LAB” configuration

| Parameter | Value | Unit |
|---|---|---|
| $T_B$ | 50 | mK |
| $T_C$ | 99 | mK |
| | | |
| $R_0$ | 139 | mΩ |
| $R_N$ | 196 | mΩ |
| $R_P$ | 0 | mΩ |
| $R_S$ | 8.142 | mΩ |
| $T_S$ | 150 | mK |
| $\alpha_c$ | 50 | |
| $P_{Heater}$ | 0 | nW |
| | | |
| $C_{TES}$ ($T_{TES,0} = 102.5\ mK$) | 3.52 | pJ/K |
| $C_{ABS}$ ($T_{ABS,0} = 94.3\ mK$) | 43.5 | pJ/K |
| $k_{AB}$ | 5.866 | $\mu W/K^4$ |
| $n_{AB}$ | 4 | |
| $k_{TA}$ | 935 | $\mu W/K^6$ |
| $n_{TA}$ | 6 | |
| | | |
| $V_{SQ,0}$ | -50 | µV |
| $V_{SQ,1}$ | 500 | µV |
| $1/M_{IN}$ | 12.1 | µA/ $\Phi_0$ |
| $1/M_{FB}$ | 40.4 | µA/ $\Phi_0$ |
| $Rf$ | 100 | kΩ |
| $A$ | 2000 | V/V |
| $\omega_0$ | 1 | MHz |
| $\omega_1$ | 9.2 | MHz |
| $t_D$ | 10 | ns |
| $G_{WBEE}$ | -10 | V/V |
| $f_{LP}$ | 30 | kHz |
| | | |
| $T_W$ | 150 | K |
| $R_W$ | 15 | Ω |
| | | |
| $N_{SQUID}$ | 0.4 | $\mu\Phi_0$ / √Hz |
| $N_{LNA}$ | 0.33 | nV/ √Hz |
| $N_{DAQ}$ | 0.48 | µV/ √Hz |

## 5. CRYOAC TRIGGER ALGORITHM AND VETO STRATEGY

In this section we present the CryoAC trigger algorithm and the veto strategy.

### 5.1 The CryoAC trigger algorithm

The simulated detector signals can be analyzed by the trigger algorithm developed for the CryoAC Warm Back End Electronics (WBEE) (see [11]), which has been integrated into the end-to-end simulator. It is based on 6 average moving windows (Fig. 9) and has 6 programmable parameters.

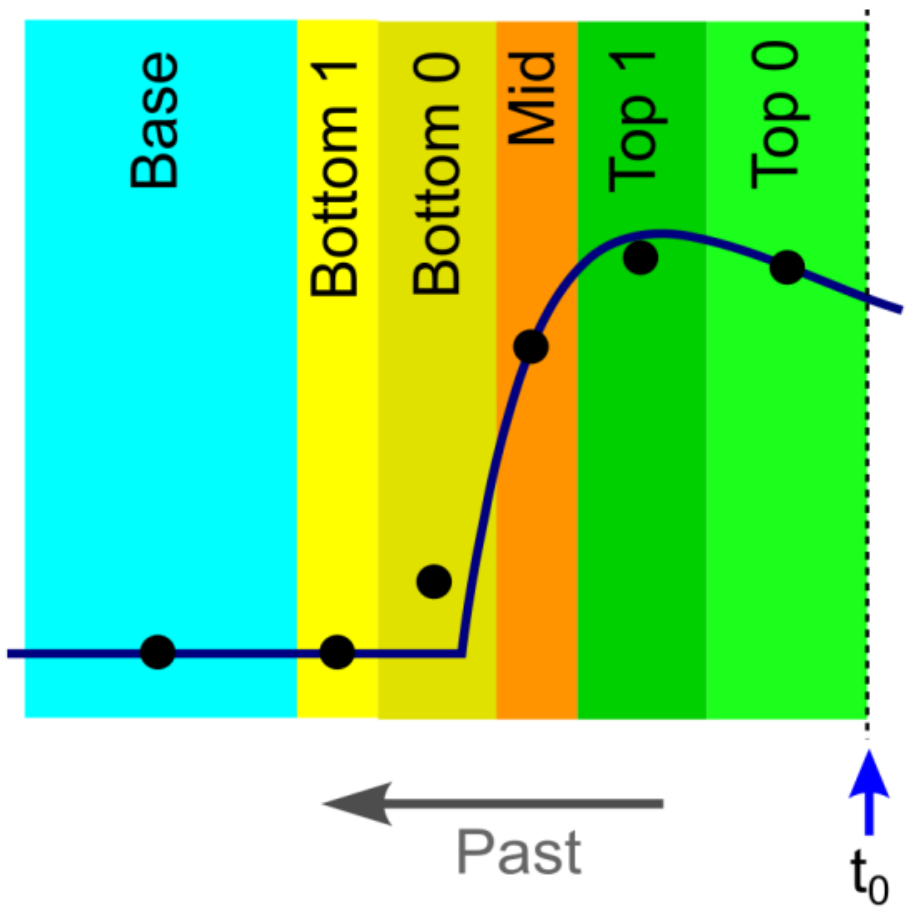

Figure 9. Sequence and naming of the average moving windows used by the trigger algorithm, with respect to the time arrow.

The moving windows are mainly used to evaluate the signal level ($y$) and quantities proportional to its first ($dy$) and second ($d^2y$) derivatives:

$$y_i = Mean(Mid) \tag{29}$$

$$dy_i = Mean\ (Top\ 1) - Mean\ (Mid) \tag{30}$$

$$d^2y_i = Mean\ (Top\ 1) - 2 \cdot Mean(Mid) + Mean\ (Bottom\ 0) \tag{31}$$

The windows length ($t_{win}$) is a programmable parameter. Trigger conditions are:

$$dy_i > Thr1 \tag{32}$$

$$d^2y_i > Thr2 \tag{33}$$

$$d^2y_{i-1} < d^2y_i < d^2y_{i+1} \tag{34}$$

$$Time\ from\ last\ trigger > t_{blind} \tag{35}$$

$$Saturation\ Status = FALSE \tag{36}$$

where $Thr1$, $Thr2$ and $t_{blind}$ are programmable parameters. The saturation status transitions are defined as:

$$Saturation\ Status\ form\ FALSE\ to\ TRUE \qquad when \qquad y_i > Sat_Thr_IN \tag{37}$$

$$Saturation\ Status\ form\ TRUE\ to\ FALSE \qquad when \qquad y_i < Sat_Thr_OUT \tag{38}$$

where $Sat_Thr_IN$ and $Sat_Thr_OUT$ are programmable parameters.

The algorithm has been developed and optimized to be robust in different cases (low pulses, pule-ups, saturations, …). The set of optimal parameters used to trigger simulated CryoAC DM127 signals is reported in Tab. 3, and some examples of trigger applications is shown in Fig. 10.

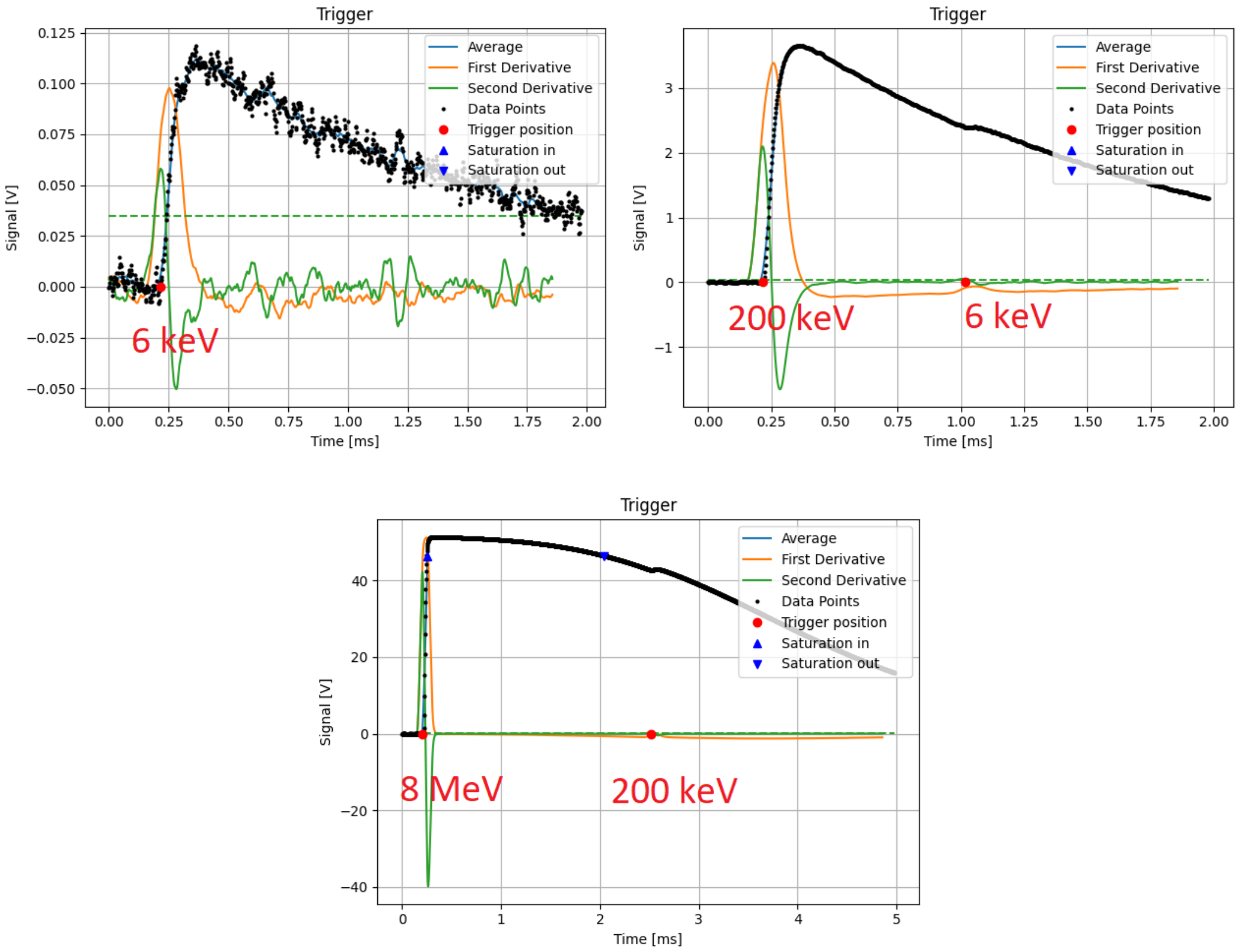

Figure 10. Application of the trigger algorithm to simulated CryoAC DM127 signals. Black points represent data, blue line the averaged signal, the orange line the evaluated first derivative and the green line the evaluated second derivative. Red points highlight the trigger positions, and blue points the saturation transitions. (*Top Left*) A small pulse around the low energy threshold. (*Top Right)* A pile-up event. (*Bottom)* A pile-up event including a saturated pulse.

Table 3. Trigger parameters optimized for the CryoAC DM127

| **Parameter** | **Value** | **Comment** |
|---|---|---|
| $t_{win}$ | 50 µs | |
| $Thr1$ | -10 V | |
| $Thr2$ | 0.035 V | ~5 σ NOISE |
| $Sat_Thr_IN$ | 46.5 V | 90% saturation level |
| $Sat_Thr_OUT$ | 46.5 V | 90% saturation level |
| $t_{blind}$ | 26 µs | |

## 5.2 The veto strategy

Each trigger on the CryoAC is associated with a veto window. Any particle detected by the TES array within this window is effectively marked as a potential background event and discarded. For non-saturated pulses the veto window starts a fixed time $\Delta t_{PRE}$ before the trigger, and it ends at a fixed time $\Delta t_{POST}$ after the trigger (Fig. 11 - left). For saturated pulses it starts at a fixed time $\Delta t_{PRE}$ before the trigger, and ends at the end of the saturation, after a time $\Delta t_{SAT}$ depending on the saturated pulse energy (Fig. 11 – right). The XIFU CryoAC requirements define the minimum veto windows acceptable by the system:

$$\Delta t_{PRE} > 26\ \mu s \tag{39}$$

$$\Delta t_{POST} > 26\ \mu s \tag{40}$$

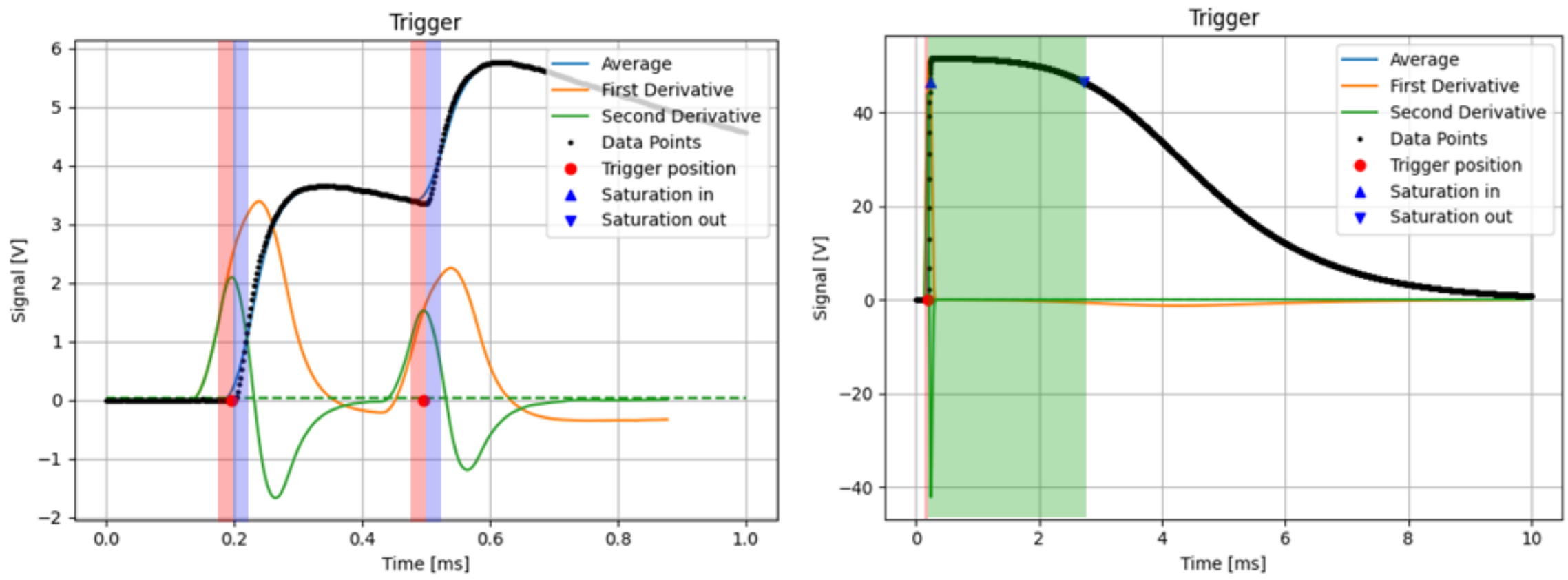

Figure 11. Veto windows overlapped to CryoAC triggered pulses. Red areas are the pre-trigger windows (length $\Delta t_{PRE} = 26\ \mu s$), blue areas the post-trigger windows (length $\Delta t_{POST} = 26\ \mu s$) and green areas the saturation windows $\Delta t_{SAT}$). *(Left)* A pile-up event. *(Right)* A saturated pulse event.

In the context of the simulations, we consider a particle "missed" if its arrival time in the event list, summed to a random time of 5 μs RMS, is outside all the veto windows defined after the triggering. The random time addition is to deal with the error in the time reference calibrations on the XIFU instrument.

The CryoAC dead time is then defined as the sum of all the veto windows durations.

Finally, the time tagging accuracy of each event is the difference between the arrival time and the nearest trigger time.

## 6. FAST SIMULATION STRATEGY

The “full” CryoAC simulator is quite time consuming (about 5 ks needed on a standard desktop PC to simulate 1 s of data stream) and therefore not usable to simulate the full data stream corresponding to the event list (i.e. about 1 Gs needed to simulate 190.8 ks of observation time). This is due to the small time step required to simulate the SQUID FLL dynamics (typically ΔX = 2ns) and properly handle the cable delay (order of tens of ns).

We have then developed a “fast” version of the simulator and defined an optimized simulation strategy to reduce the computation time, without losing accuracy.

### 6.1 Fast version of the simulator

The “fast” version of the simulator does not solve the SQUID FLL equations, but it describes the SQUID system as a Global Gain $G_{SQUID\ FLL}$ and a first order low-pass filter with cutoff frequency $f_{LP\ FLL}$. The output of the FLL is then conditioned as usual by the WBEE:

$$V_{FLL\ OUT\ (i+1)} = \left(I_{TES\ (i)} - I_{TES\ (0)}\right) \cdot \alpha_{SQUID\ FLL}\ G_{SQUID\ FLL} + V_{FLL\ OUT\ (i)} \cdot \left(1 - \alpha_{SQUID\ FLL}\right) + N_{FLL} \tag{41}$$

$$G_{SQUID\ FLL} = R_F \cdot M_{IN}/M_{FB} \tag{42}$$

$$\alpha_{SQUID\ FLL} = dX/[1/(2\pi f_{LP\ FLL}) + dX] \tag{43}$$

$$V_{OUT\ (i+1)} = V_{FLL\ OUT\ (i+1)} \cdot \alpha_{WBEE}\ G_{WBEE} + V_{OUT\ (i)} \cdot (1 - \alpha_{WBEE}) + N_{DAQ} \tag{44}$$

where the noise $N_{FLL}$ is manually defined in order to reproduce the total noise of the “full” (electrothermal + SQUID FLL) simulations. Fig. 12 shows a comparison between the “full” and the “fast” simulations output, which are fully compatible.

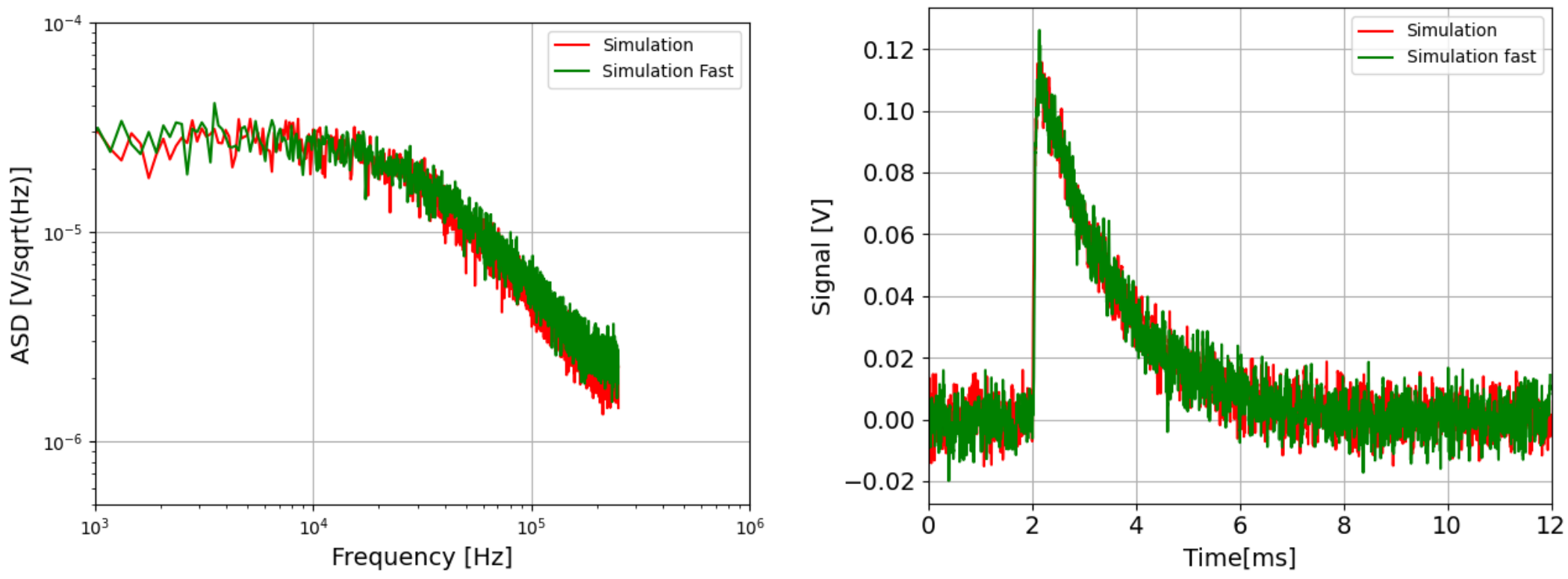

Figure 12. Comparison between the ”fast” (green lines) and the “full” (red lines) simulations output. *(Left)* Simulated noise power spectrum. *(Right)* Simulated 6 keV pulse.

The use of the fast simulator allows to strongly increase the simulation step $dX$ (typically from 2 ns to 2 µs), thus reducing simulation time, without losing accuracy. The strategy is then to use the “full” simulator to properly evaluate the noise and check the stability of the system at different energies (up to the saturation regime and the maximum energies in the event list), and then use the “fast” simulator for the bulk simulations.

### 6.2 Optimized simulation strategy

Even using the “fast” simulator, simulating the full CryoAC data stream corresponding to the event list is fairly time consuming (around 3.5 s needed on a standard desktop PC to simulate 1 s of data stream, corresponding to about 670 ks to simulate 190.8 of ks observation time).

We have therefore defined an optimized simulation strategy, to be used in conjunction with the fast simulator. Given the event list, instead of simulate the entire datastream, we can divide the events into different categories and, depending on the category, simulate only a relevant part of the event. The categories are:

- **Single events**. They are events that are not followed by other events within a $t^*$ time. In this case we simulate just the rise of the pulse (from a time $t_{PRE}$ before the arrival time to a time $t_{POST}$ after the arrival time), in order to check triggering and time tagging accuracy (Fig. 13 top).
- **Pile-up events.** They are events followed by other events within a $t^*$ time. In this case we simulate the entire signal stream between pulses to properly test the trigger efficiency in this condition (Fig. 13 middle).
- **Saturated events.** They are events that reach saturation. In this case we simulate the entire event (until the return in baseline) to proper evaluate dead time (Fig. 13 bottom).

In addition, we also simulate some noise stream (i.e. without events) to evaluate the false-trigger rate.

With this strategy, given $t^*$ = 10 ms, $t_{PRE}$ = 200 μs and $t_{POST}$ = 200 μs, we are able to perform a full simulation of the event list (571’135 events, corresponding to 190.8 ks of observation time) in ~ 4 hours on a standard desktop PC.

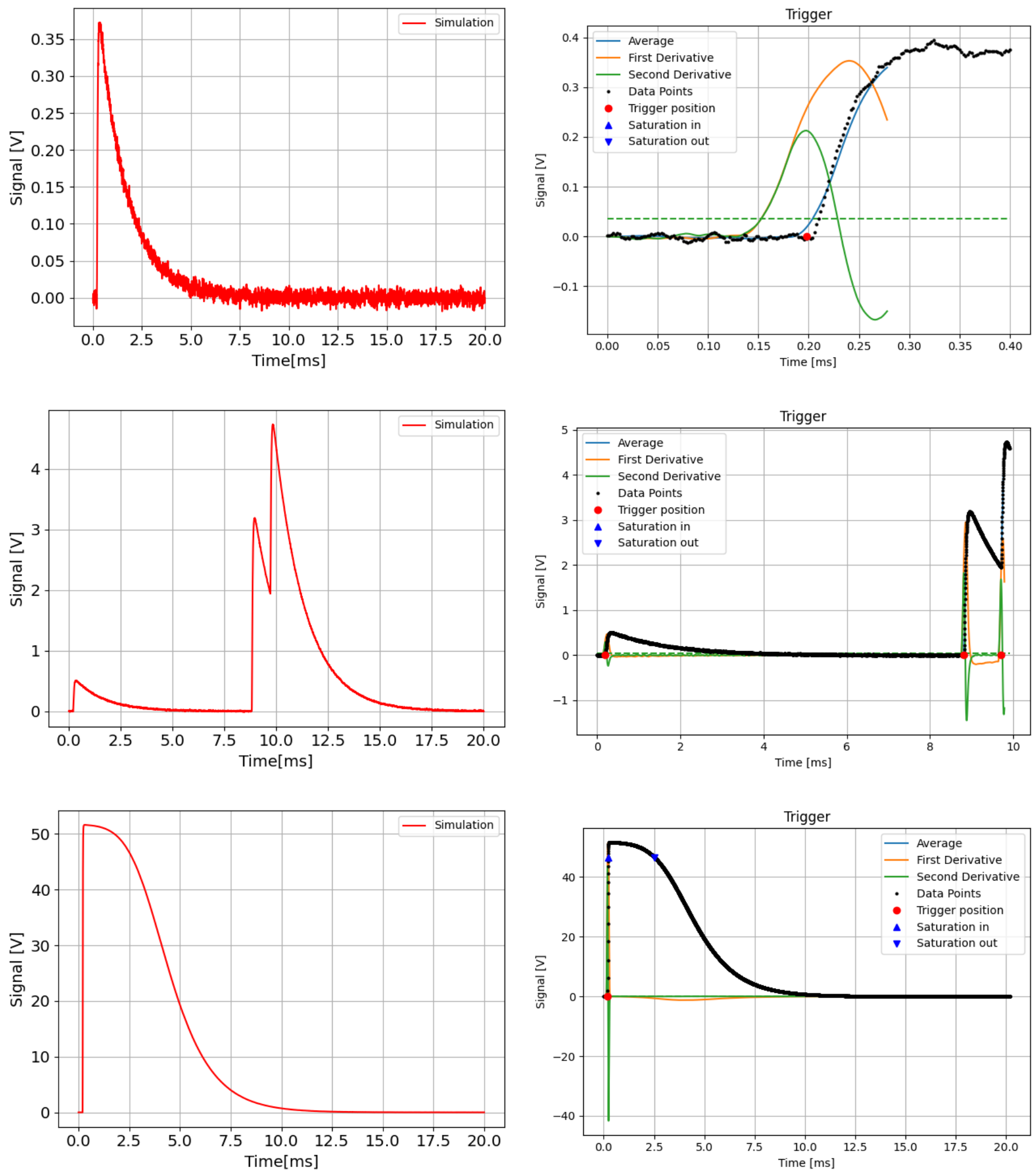

Figure 13. Application of the optimized simulation strategy for events of different categories. The full events are shown on the *Left*, while the relevant parts to be simulated and analyzed are shown on the *Right*. *(Top)* A single event. *(Middle)* A pile-up event. *(Bottom)* A saturated event.

# 7. SIMULATION RESULTS

Finally, in this section we report the results of the simulations. We started by simulating the DM127 prototype as an on-flight pixel, and then defined and tested different CryoAC pixel configurations in the new Athena context.

## 7.1 Lab prototype DM127 as on-flight pixel

To have a first reference, we have initially performed a full study simulating the DM127 prototype as on-flight pixel in the L1 environment. The simulator setup is the one shown in Tab. 2. The event list has been modified considering a detector area of 1 $cm^2$, resulting in a count rate of 4.0 cts/s/pixel. The results of the simulation are reported in Tab. 4, and compared with the CryoAC single-pixel requirements.

Table 4. DM127 as on-flight pixel simulation results.

| Configuration | Veto strategy | Dead Time | Missed Particles |
|---|---|---|---|
| *Requirements* | - | *0.25 %* | *0.014% (76/544994)* |
| DM127 (1 $cm^2$) LAB | Minumum veto windows ($\Delta t_{POST}$ = 26 μs) | 0.03 % ✓ | 0.025 % (136/544994) X |
| DM127 (1 $cm^2$) LAB | Optimized veto windows ($\Delta t_{POST}$ = 200 μs) | 0.11 % ✓ | 0.004 % (21/544994) ✓ |

Using the minimum acceptable veto windows shown in Sec. 5.2 (i.e. $\Delta t_{PRE}$ = 26 μs, $\Delta t_{POST}$ = 26 μs), we obtained a deadtime 8 times lower than the requirement, but a number of missed particles about a factor of 2 higher. These missed events are mainly pile-up events over a pulse rise (see Fig. 14 - left).

Thus, we decided to adopt a different veto strategy by increasing the veto windows after the trigger, i.e. $\Delta t_{POST}$ = 200 μs (see Fig. 14 - right). This means to strongly reduce the possibility to miss an event, at the cost of an increased deadtime. With this new veto strategy, we obtained a result that fulfils both requirements, with more than a factor of 2 margin (see Tab. 4).

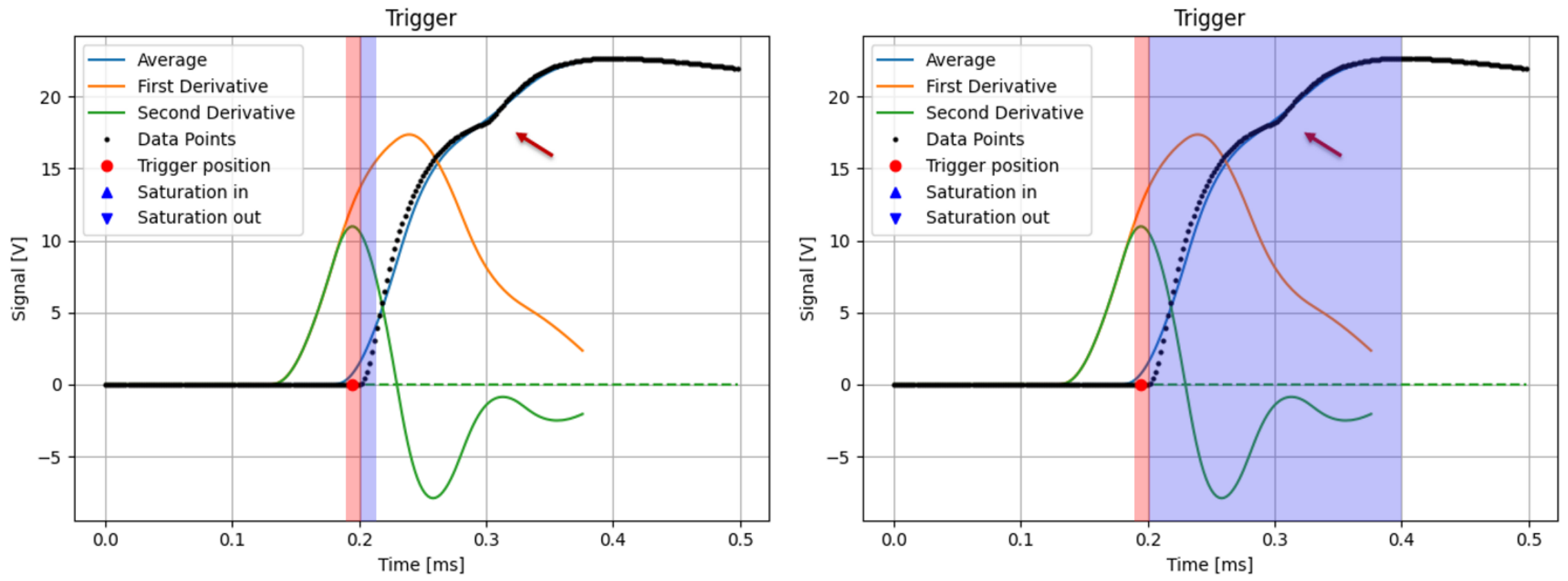

Figure 14. – Typical pileup event missed by the minimum deadtime veto strategy. *(Left)* The second particle arriving ~ 100 μs after the first event (red arrow) is not triggered. Being outside the veto windows (shaded zones) the event is missed. *(Right)* The Same event shown with the optimized veto windows ($\Delta t_{POST}$ = 200 μs). In this case the missed pulse is inside the veto window, thus it is not missed.

Finally, Fig. 15 shows the time tagging accuracy distribution of the simulated events. Most of the events are triggered within 4 μs from the arrival time, largely in-line with requirement (10 μs at 1σ).

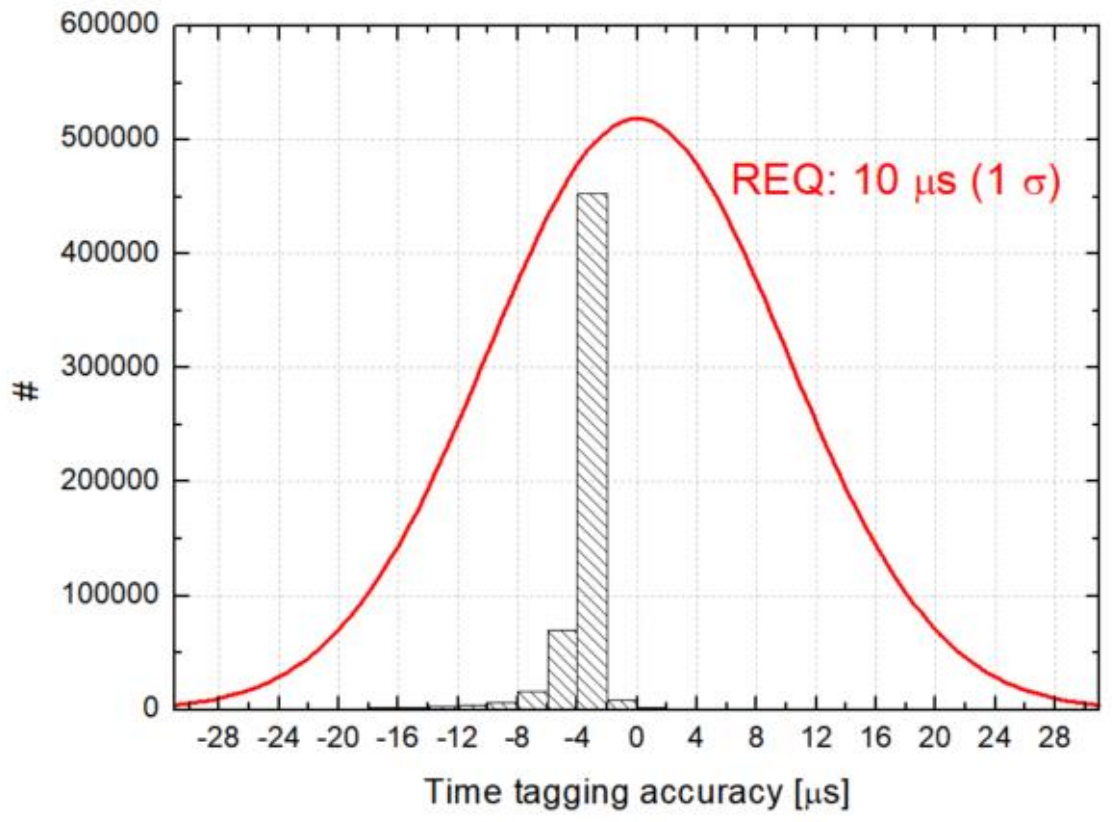

Figure 15. Time tagging accuracy distribution.

### 7.2 New ATHENA configurations

Following the first results, we have simulated two different CryoAC pixel configurations in the new Athena framework.

The first configuration, codename "DM127 (0.75 $cm^2$) new Athena", has been defined starting from the DM127 prototype characteristics. The detector area has been reduced to 0.75 $cm^2$, the cable delay and noise have been redefined considering a 4.5 m physical distance between the cold stage and the warm electronics (with respect to the 1.5 m in the lab setup), and the LNA noise have been redefined following the flight electronics requirement. Finally, The SQUID characteristics have been modified to approximate the shape of the last CryoAC SQUID prototype (model O4 by VTT), which featured well-shaped characteristics (see Fig. 16). The full set of parameters used in this simulation is reported in Tab. 5.

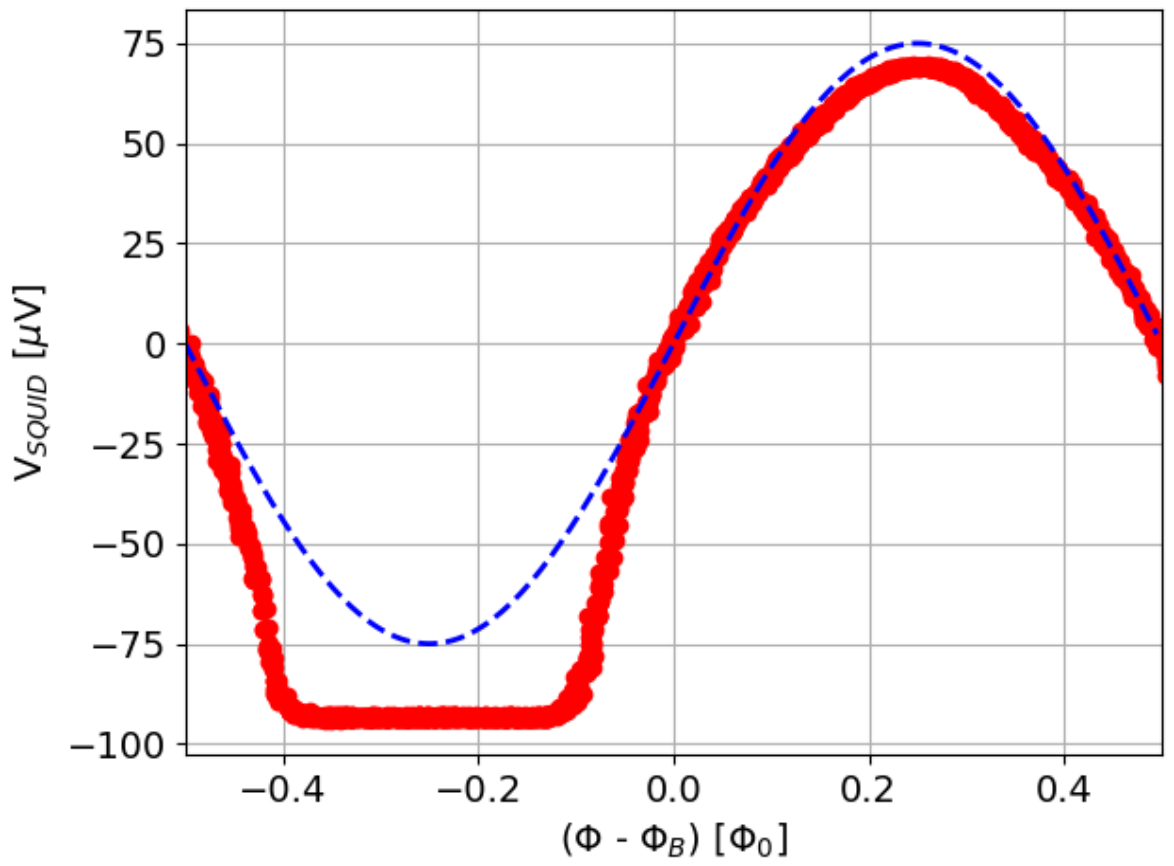

Figure 16. The V-Φ characteristic curve of the last CryoAC SQUID prototype (VTT model O4), the new reference for the simulations. The red points are real data, while the blue dashed line shows the V-Φ sinusoidal representation inside the simulator.

Table 5. "DM127 (0.75 cm$^2$) new Athena" configuration. The parameters modified with respect to the "DM127 (1 cm$^2$) LAB" configuration are shown in blue.

| Parameter | Value | Unit |
|---|---|---|
| $T_B$ | 50 | mK |
| $T_C$ | 99 | mK |
| | | |
| $R_0$ | 139 | mΩ |
| $R_N$ | 196 | mΩ |
| $R_P$ | 0 | mΩ |
| $R_S$ | 8.142 | mΩ |
| $T_S$ | 150 | mK |
| $\alpha_c$ | 50 | |
| $P_{Heater}$ | 0 | nW |
| | | |
| **$C_{TES}$ ($T_{TES,0}$ = 102.3 mK)** | **3.51** | **pJ/K** |
| **$C_{ABS}$ ($T_{ABS,0}$ = 94.0 mK)** | **33.7** | **pJ/K** |
| $k_{AB}$ | 5.866 | µW/K$^4$ |
| $n_{AB}$ | 4 | |
| $k_{TA}$ | 935 | µW/K$^6$ |
| $n_{TA}$ | 6 | |
| | | |
| **$V_{SQ,0}$** | **0** | **µV** |
| **$V_{SQ,1}$** | **75** | **µV** |
| **$1/M_{IN}$** | **12.4** | **µA/ $\Phi_0$** |
| **$1/M_{FB}$** | **20.1** | **µA/ $\Phi_0$** |
| $Rf$ | 100 | kΩ |
| $A$ | 2000 | V/V |
| $\omega_0$ | 1 | MHz |
| $\omega_1$ | 9.2 | MHz |
| **$t_D$** | **30** | **ns** |
| $G_{WBEE}$ | -10 | V/V |
| $f_{LP}$ | 30 | kHz |
| | | |
| $T_W$ | 150 | K |
| **$R_W$** | **70** | **Ω** |
| | | |
| $N_{SQUID}$ | 0.4 | µ$\Phi_0$ / √Hz |
| **$N_{LNA}$** | **1.0** | **nV/ √Hz** |
| $N_{DAQ}$ | 0.48 | µV/ √Hz |

The second configuration, codename “Opt pixel (0.75 $cm^2$) new Athena”, finally includes a factor 3 higher thermal conductance from the absorber to the thermal bath (as foreseen in the next prototype CryoAC DM1.1) and an optimized TES working point, which was not available on DM127 due to the transition issues (see Fig. 17). The full set of parameters used in this simulation is reported in Tab. 6.

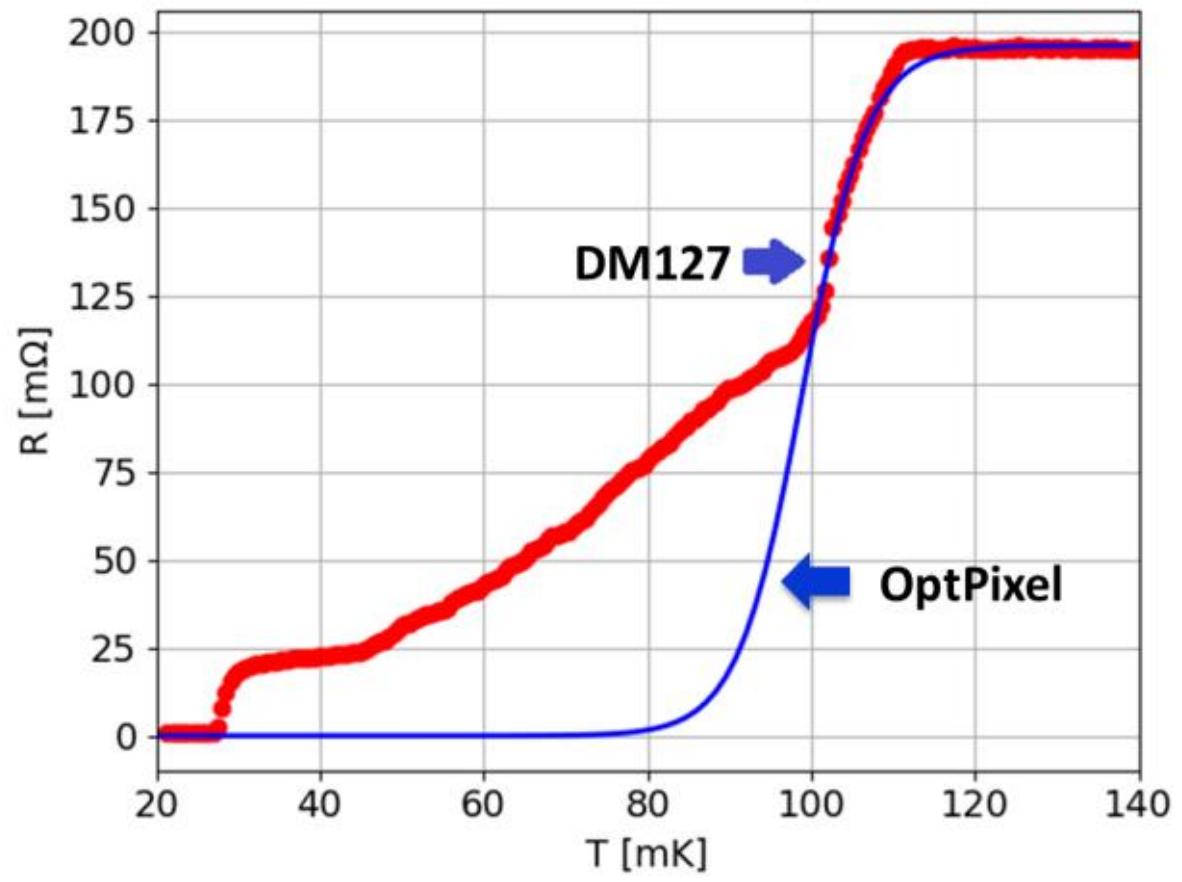

Figure 17. The CryoAC DM127 superconducting transition (red points) and its description inside the simulator (blue line). Blue arrows highlights the TES working points in DM127 and Opt Pixel simulator configurations.

Table 6. “Opt Pixel (0.75 cm$^2$) new Athena” configuration. The parameters modified with respect to the “DM127 (1 cm$^2$) LAB” configuration are shown in blue, and the parameters modified with respect to the “DM127 (0.75 cm$^2$) new Athena” configuration are shown in red.

| Parameter | Value | Unit |
|---|---|---|
| $T_B$ | 50 | mK |
| $T_C$ | 99 | mK |
| | | |
| **$R_0$** | **40** | **mΩ** |
| $R_N$ | 196 | mΩ |
| $R_P$ | 0 | mΩ |
| $R_S$ | 8.142 | mΩ |
| $T_S$ | 150 | mK |
| $\alpha_c$ | 50 | |
| $P_{Heater}$ | 0 | nW |
| | | |
| **$C_{TES}$ ($T_{TES,0}$ = 93.5 mK)** | **3.21** | **pJ/K** |
| **$C_{ABS}$ ($T_{ABS,0}$ = 75.2 mK)** | **17.3** | **pJ/K** |
| **$k_{AB}$** | **17.6** | **µW/K$^4$** |
| $n_{AB}$ | 4 | |
| $k_{TA}$ | 935 | µW/K$^6$ |
| $n_{TA}$ | 6 | |
| | | |
| **$V_{SQ,0}$** | **0** | **µV** |
| **$V_{SQ,1}$** | **75** | **µV** |
| **$1/M_{IN}$** | **12.4** | **µA/ $\Phi_0$** |
| **$1/M_{FB}$** | **20.1** | **µA/ $\Phi_0$** |
| $Rf$ | 100 | kΩ |
| $A$ | 2000 | V/V |
| $\omega_0$ | 1 | MHz |
| $\omega_1$ | 9.2 | MHz |
| **$t_D$** | **30** | **ns** |
| $G_{WBEE}$ | -10 | V/V |
| $f_{LP}$ | 30 | kHz |
| | | |
| $T_W$ | 150 | K |
| **$R_W$** | **70** | **Ω** |
| | | |
| $N_{SQUID}$ | 0.4 | µ$\Phi_0$ / √Hz |
| **$N_{LNA}$** | **1.0** | **nV/ √Hz** |
| $N_{DAQ}$ | 0.48 | µV/ √Hz |

The simulation results are reported in Tab. 7, and compared with the old "DM127 (1 $cm^2$) LAB" simulation and the CryoAC single pixel requirements. New Athena simulations have been performed by using the event list shown in Sec. 2 (i.e. count rate of 3.0 cts/s/pixel) and the optimized veto strategy defined in Sec. 7.1 (i.e. veto window $\Delta t_{POST}$ = 200 μs). In both cases, the detector is within its requirements, with a margin of more than a factor 2 for both dead time and missed particles. Furthermore, all simulated configurations also meet the FPA power dissipation and bias current constraints. Finally, we report that the time tagging accuracy is always within the requirement with a large margin, with distribution similar to the one shown in Fig. 15.

Table 7. Simulation results for the CryoAC new Athena configurations.

| **Configuration** | **Power dissipation** | **Bias current** | **Dead Time** | **Dead Time in saturation** | **Missed Particles** | | |
|---|---|---|---|---|---|---|---|
| *Requirements* | *10 nW* | *3000 μA (1000 μA goal)* | *0.25 %* | - | *0.014%* | *(76/544994)* | |
| | | | | | | | |
| DM127 (1 $cm^2$) LAB | 9.3 nW | 1000 μA | 0.11 % | 0.02 % | 0.004 % | (21/544994) | ✓ |
| | | | | | | | |
| **DM127 (0.75 $cm^2$) new Athena** | **8.4 nW** | **1000 μA** | **0.08 %** | **0.02 %** | **0.004 %** | **(21/544994)** | ✓ |
| **Opt Pixel (0.75 $cm^2$) new Athena** | **4.2 nW** | **630 μA** | **0.08 %** | **0.02 %** | **0.003 %** | **(18/544994)** | ✓ |

## 8. CONCLUSIONS

We have presented the end-to-end simulator of the ATHENA X-IFU Cryogenic Anticoincidence Detector (CryoAC). It has been developed to drive the detector design and verify the compliance with its requirements. The simulator has been tuned, validated and tested with the last single-pixel CryoAC prototype, namely DM127. Finally, it has been used to test different pixel configurations in the new Athena frameworks. The results are fully encouraging, and show that there are realistic in-flight CryoAC configurations that can meet all the requirements with a good margin. The next step will now be to perform a full set of simulations moving the detector parameters one by one, in order to assess the sensitivity of the results and to precisely define the parameter space within which the CryoAC detector can be designed.

## ACKNOWLEDGEMENTS

This work has been supported by ASI (Italian Space Agency) contract n. 2019-27-HH.0. The authors would like to acknowledge VTT for supplying the CryoAC SQUID prototypes.